# Minimizing The Expected Distortion Using Multi-layer Coding Approach In Two-Hop Networks

Sayed Ali Khodam Hoseini, Soroush Akhlaghi and Mina Baghani

Department of Engineering,Shahed University, Tehran, Iran;

Engineering Department, Iran

Emails: {khodamhoseini, akhlaghi }@shahed.ac.ir; baghani@aut.ac.ir

## Abstract

This work concerns minimizing the achievable distortion of a Gaussian source over a two-hop block fading channel under mean square-error criterion. It is assumed that there is not a direct link between transmission ends and the communication is carried out through the use of a relay, employing the Decode and Forward (DF) strategy. Moreover, for both of transmitter-to-relay and relay-to-destination links the channel statistics are merely available at the corresponding receivers, while the channel gains are not available at the affiliated transmitters. It is assumed that a Gaussian source is hierarchically encoded through the use of successive refinement source coding approach and sent to the relay using a multi-layer channel code. Similarly, the relay transmits the retrieved information to the destination through the use of a multi-layer code. Accordingly, in a Rayleigh block fading environment, the optimal power allocation across code layers is derived. Numerical results demonstrate that the achievable distortion of the proposed multi-layer approach outperforms that of single-layer code. Moreover, the resulting distortion of DF strategy is better than Amplify and Forward (AF) strategy for channel mismatch factors greater than one, while the resulting distortion has a marginal degradation for channel mismatch factors lower than one.

## I. INTRODUCTION

Relaying is mostly regarded as a promising approach to enhance the quality and coverage of wireless communication networks. In this regard, various coding strategies are deployed at the relays, among them, the Amplify-and-Forward (AF) and the Decode-and-Forward (DF) strategies are mostly addressed in the literature. Note that there is not a single coding strategy which performs well for all network topologies and channel realizations. For instance, in a two-hop channel when the direct link between the source and the destination is in poor condition and the destination has only access to the information exchanged by the relay, it is demonstrated that the DF is optimal in terms of maximizing the achievable rate [1]. This motivated us to study the performance of such a relaying scheme when dealing with distortion and compare the result to the AF strategy.

We consider a point-to-point two-hop communication channel, where there is not a direct link between transmission ends. The channel is assumed to be constant throughout one transmission block and varies independently for the next blocks. Moreover, the relay is assumed to be simple, meaning it cannot do buffering, water-filling across time or coding over consecutive blocks. Also, it is assumed that the channel gain of each hop is not available at the corresponding transmitter.

It is widely recognized that when the channel gain is not available at the transmitter, the so-called broadcast strategy, incorporating multi-layer coding is optimal in terms of maximizing the average achievable rate [2], [3]. This problem has been extended to relay networks and the optimality of using such an approach is studied in various cases [4]–[6].

On the other hand, the successive refinement (SR) source coding approach is useful when trying to broadcast the source information to multiple receivers with different channel conditions. In the SR method, the source information is hierarchically refined through multiple source layers such that the upper layers are refinements of lower source layers and the base layer is the most protected one.

According to the notion of broadcast approach, a point-to-point communication link in the lack of CSIT can be modeled by a point to multi-point network with an infinite number of virtually ordered receivers, each corresponding to a channel realization. This motivated Tian *et al.* to concurrently incorporate the SR source coding and the broadcast approaches to derive an achievable distortion of Gaussian source over a single hop channel when the CSIT is not available [7]. This approach is further studied in a two-hop relay-assisted network when the relay knows the channel information associated with both hops, thus a single layer code is incorporated at the second hop [8], [9].

What differentiates the current study to the works done in [8], [9] is that the channel gain associated with the second hop is not available at the relay, thus the use of single layer code at the relay is not an appropriate choice. The main contribution of the current study is to find the best power allocation policies associated with both hops leading to the minimum expected distortion at the destination. This is found to be a computationally infeasible problem. However, as is shown later, using some approximations, an analytical close to optimal solution is derived.

The reminder of this paper is organized as follows. The background information regarding the use of multi-layer source coding approach in a single hop channel is presented in Section II. Section III extends the terminology to a two-hop network and presents the results of Rayleigh fading channel. Finally, Section V summarizes findings.

## II. BACKGROUND INFORMATION

In a point-to-point block fading channel, the $k^{th}$ received signal at the destination for a given transmission block can be written as,

$$y_k = hx_k + n_k, \tag{1}$$

where $x_k$ and $y_k$ denote respectively, the complex channel input and the corresponding output at the aforementioned transmission block. $n_k$ is circularly symmetric zero-mean complex Gaussian noise with unit variance, i.e., $n_k \sim \mathcal{CN}(0,1)$. Moreover, $h \in \mathbb{C}$ represents the channel fading coefficient of current transmission block with strength $\gamma = |h|^2$, which is assumed to be constant throughout each transmission block and varies independently for the next blocks.

Let's assume the source symbols of length $l_s$ are sent by the use of channel codewords of length $l_c$. In this case, the source/channel mismatch factor is computed as $l_c/l_s$. Moreover, the channel coding blocks are assumed to be long enough to approach the rate-distortion limit, while still shorter than the coherence time of the channel.

It is assumed that the transmitter makes use of a multi-layer code with infinite number of layers, in which the fractional power $\rho(\gamma)d\gamma$ is set to the layer indexed by $\gamma$. Thus, the corresponding fractional rate becomes

$$dR(\gamma) = \log\left(1 + \frac{\gamma\rho(\gamma)d\gamma}{1+\gamma T(\gamma)}\right) \simeq \frac{\gamma\rho(\gamma)d\gamma}{1+\gamma T(\gamma)}. \tag{2}$$

In (2), $T(\gamma)$ is the power assigned to the layers with channel strengths above $\gamma$, which are not decodable at the destination when the channel strength is $\gamma$, i.e.,

$$T(\gamma) = \int_\gamma^\infty \rho(u)du. \tag{3}$$

Consequently, the total achievable rate decoded at the destination with channel strength $\gamma$ is the integration of all fractional rates allocated to layers below $\gamma$, i.e.,

$$R(\gamma) = \int_0^\gamma \frac{u\rho(u)du}{1 + uT(u)}. \tag{4}$$

On the other hand, the distortion of a Gaussian source transmitted to the destination when the channel strength is $\gamma$, i.e., $D(\gamma)$, can be written as,

$$D(\gamma) = \exp\big(-bR(\gamma)\big). \tag{5}$$

To formulate the power constraint and the average achievable distortion at the destination, an auxiliary function is defined in [7] as follows,

$$I(\gamma) = \exp\big(R(\gamma)\big) = D(\gamma)^{\frac{-1}{b}}. \tag{6}$$

Using (6), the minimum average achievable distortion when the transmit power constraint is subjected to $P_t$, can be written as,

$$\overline{D}(I) = \min_{I(.)} \int_0^\infty \frac{f(\gamma)}{I(\gamma)^b} d\gamma$$
$$\text{s.t. } P(I) = \int_0^\infty \frac{I(\gamma)}{\gamma^2} d\gamma \leq P_t, \tag{7}$$

where $f(\gamma)$ denotes the probability density function (pdf) of channel strength $\gamma$. Accordingly, in [7] the best function $I^{opt}(\gamma)$ minimizing (7) is derived, showing the optimal function should have at most one continues single interval in the region which meets the inequality,

$$\frac{d}{d\gamma}\big(\gamma^2 f(\gamma)\big) > 0. \tag{8}$$

For instance, in Rayleigh channel it is shown that there is just one continuous interval like $\gamma \in [\gamma_1, \gamma_2]$ where (8) is satisfied. Moreover, the values of $\gamma_1$ and $\gamma_2$ as well as the function $I^{opt}(\gamma)$ are analytically derived in [7].

## III. DISTORTION MINIMIZATION FOR DF RELAYING

This section tends to minimize the expected achievable distortion of a Gaussian source at the destination of a two-hop network. It is assumed that the transmitter and the relay have no access to the corresponding transmit channel strengths, thus these nodes make use of successive refinement source coding and multi-layer channel coding for transmitting information to the associated receivers. Therefore, the problem is to find the optimum power allocation functions across code layers associated with the source and relay nodes, leading to the minimum average achievable distortion at the destination. To this end, the problem is divided into two main parts. First, knowing the available distortion at the relay, the optimal transmission strategy at this node is derived. Then, having the strategy of the second hop, the optimal power allocation strategy at the fist hop is calculated.

*A. The Optimal Power Allocation Strategy At the Relay*

Assuming the instantaneous channel strength associated with the first hop is $\gamma$, the relay can successfully decode code layers assigned to channel strengths lower than $\gamma$, thus it can decode the source information at some distortion level, i.e., $D_r(\gamma)$. Accordingly, the problem is first formulated as an optimization problem, where it is argued that the underlaying problem is convex and more importantly, as it will be proved later on this section, the optimal solution has a single non-zero power allocation interval.

To this end, assuming the second hop's channel strength is denoted by $l$ and considering a single non-zero power allocation interval $l \in [l_1, l_2]$, the average achievable distortion at the destination becomes [8],

$$\overline{D}(I_r|D_r(\gamma)) = \int_0^\infty \frac{f_r(l)}{I_r(l)^b} dl = \int_{l_1}^{l_2} \frac{f_r(l)}{I_r(l)^b} dl \\ + F_r(l_1) + \frac{1 - F_r(l_2)}{I_r(l_2)^b}, \tag{9}$$

where $I_r(l)$ is the second hop's auxiliary function defined as follows,

$$I_r(l) = \exp\big(R_r(l)\big) \tag{10}$$

On the other hand, assuming the relay is subject to the power constraint $P_r$, it follows,

$$P(I_r|D_r(\gamma)) = \int_0^\infty \frac{I_r(l)}{l^2} dl \\ = \int_{l_1}^{l_2} \frac{I_r(l)}{l^2} dl + \frac{I_r(l_2)}{l_2} - \frac{1}{l_1} \leq P_r. \tag{11}$$

In (9) and (11), the input relay's distortion, i.e., $D_r(\gamma)$, is a function of the source auxiliary function, i.e., $D_r(\gamma) = I_t(\gamma)^{-b} = \exp\big(-bR_t(\gamma)\big)$. By the same token, the resulting distortion at the destination can be expressed as $D_d(l) = I_r(l)^{-b}$.

It is worth mentioning that as the destination's received signal is a degraded version of the received signal at the relay. Thus, the instantaneous distortion at the destination can not be smaller than that of the relay's. As a result, considering there is just a single power allocation interval, and noting the best achievable distortion at the destination is the same as that of available at the relay, it follows,

$$D_d(l_2) = \frac{1}{I_r(l_2)^b} \geq D_r(\gamma). \tag{12}$$

Moreover, noting the power allocation function is non-negative (rate allocation function is positive) and referring to (10), it follows,

$$\frac{d}{dl}I_r(l) > 0. \tag{13}$$

Incorporating the variational form of the integrands of (9) and (11) as $\mathcal{D}(I_r, I'_r, l) = \frac{f_r(l)}{I_r(l)^b}$ and $\mathcal{P}(I_r, I'_r, l) = \frac{I'_r(l)}{l^2}$, the complete form of the optimization problem in (9) with constraints (11), (12) and (13) can be encapsulated as follows,

$$\overline{D}(I_r|D_r(\gamma)) = \min_{I_r(.)} \int_0^\infty \mathcal{D}(I_r, I'_r, l)dl$$
$$\text{s.t.} \begin{cases} \int_0^\infty \mathcal{P}(I_r, I'_r, l)dl \leq P_r \\ I'_r(l) \geq 0 \\ \frac{1}{I_r(l_2)^b} \geq D_r(\gamma). \end{cases} \tag{14}$$

To find the optimal solution, we have relaxed the second constraint, i.e., $I'_r(l) \geq 0$. However, it will be shown that the obtained result meets this constraint, thus it would be the optimal solution of (14) as well. Also, the last constraint, i.e., $\frac{1}{I_r(l_2)^b} \geq D_r(\gamma)$, is a boundary condition which merely affects the unknown parameters involved in the optimal solution. Thus, using the method of lagrange multipliers, one can fuse the objective function of (14) and the power constraint (the first constraint) into the following single-letter formulation,

$$\mathcal{L}(I_r) = \overline{D}(I_r|D_r(\gamma)) + \lambda(\mathcal{P}(I_r, I'_r, l) - P_r). \tag{15}$$

It should be noted that $\mathcal{D}(I_r, I'_r, l)$ and $\mathcal{P}(I_r, I'_r, l)$ are convex w.r.t. $I_r(.)$. Thus, adding an arbitrary non-negative function like $\delta I_r(l)$ to $I_r(l)$ and equating the linear part of the incremental function $\Delta(\delta I_r) = \mathcal{L}(I_r + \delta I_r) - \mathcal{L}(I_r)$ to zero, and noting $\delta I_r(l) \neq 0$ for $l \in (l_1, l_2)$, it follows [10],

$$\int_{l_1}^{l_2} \left[\mathcal{D}_{I_r} + \lambda \mathcal{P}_{I_r} - \frac{d}{dl}[\mathcal{D}_{I'_r} + \lambda \mathcal{P}_{I'_t}]\right]\delta I_r(l)d\gamma = 0 \tag{16}$$

It's worth mentioning that $I_r(l_1) = \exp(R_r(l_1)) = 1$, since the power allocation is assumed to take place over the interval $[l_1, l_2]$ and the source is assumed to be of unit power. As a result, as long as the channel strength is lower than the first layer, no layer is decoded and the distortion becomes the source power. Moreover, $I_r(l_2)$ can be computed from the boundary condition in (12). Thus in boundary points $l = l_1$ and $l = l_2$, no optimization will take place, and the arbitrary function in the boundary points should be zero. Therefore, the remaining non-integral components of (9) and (11) do not participate in (16). In this case, equating the integrand of (16) to zero, gives,

$$I_r(l) = \left(\frac{bl^2 f_r(l)}{\lambda}\right)^{\frac{1}{b+1}}, \tag{17}$$

where noting $I_r(l_1) = 1$, (17) changes to,

$$I_r(l) = \left(\frac{l^2 f_r(l)}{l_1^2 f_r(l_1)}\right)^{\frac{1}{b+1}} \tag{18}$$

Therefore, (12) and (18) yields the following,

$$\frac{1}{I_r(l)^b} = D_r(\gamma) \to l_2^2 f_r(l_2) = D_r(\gamma)^{\frac{-(b+1)}{b}} l_1^2 f_r(l_1) \tag{19}$$

Solving (11) and (19) together, gives the start and end points of the power allocation interval incorporated at the relay, i.e., $\{l_1, l_2\}$, for various relay's received distortion values ($D_r(\gamma)$) ranging from 0 to 1. Knowing $l_1$, the optimal value of $I_r(l)$ is completely characterized using (18). Finally, substituting the computed values of $l_1$ and $l_2$ as well as $I_r(l)$ into (9), the mean achievable distortion at the destination subject to a known available distortion at the relay can be derived.

Now, one should prove that the the assumption of a single non-zero power allocation interval is indeed true. To this end, Lemma (1) is provided.

**Lemma 1** *Consider the following minimization problem,*

$$\min_{I(.)} \int_0^\infty f(x) \mathcal{G}\left(\frac{1}{I(x)^b}\right) dx$$
$$s.t. \begin{cases} P(I) = \int_0^\infty \frac{I(x)}{x^2} dx \leq P \\ I'(x) \geq 0 \end{cases}, \tag{20}$$

*where $f(.)$ is an arbitrary non-negative function, $\mathcal{G}(.)$ is an increasing[1] and convex function and $I(.)^{-b}$ is a convex function. In this case, the optimal solution associated with (20) has the following properties:*

1) *The necessary condition to meet the last constraint is,*

$$\frac{d}{dx}\left(x^2 f(x)\right) > 0. \tag{21}$$

---

[1] The condition under which $\mathcal{G}(.)$ is an increasing function is provided in Appendix B.

2) *There is at most one single-interval of continuous, non-zero power allocation in each region, satisfying (21).*

*Proof:* see Appendix A. ∎

The optimization problem in (14) can be readily converted to (20) through setting $\mathcal{G}(x) = x$. For instance, let's consider the Gamma probability density function, subsuming some prominent distributions including exponential distribution, as follows,

$$f(x; \alpha, \beta) = \frac{\beta^\alpha x^{\alpha-1} e^{-\beta x}}{\Gamma(\alpha)}, \; for \; x \geq 0, \; and \; \alpha, \beta > 0, \tag{22}$$

Examining the condition (21) over (22), the following single interval is identified,

$$0 < x < \frac{1+\alpha}{\beta}. \tag{23}$$

Thus, the optimal solution in cases that we are dealing with any p.d.f. that can be represented by Gamma's family, results in a single continuous interval satisfying (23).

## B. The Optimal Power Allocation Strategy at the Source

Knowing the mean achievable distortion at the destination conditioned on the relay's received distortion, namely $\mathcal{G}(D_r(\gamma)) = \overline{D}(I_r|D_r(\gamma))$, the objective is to find the optimal power allocation policy at the first hop $(I_t(.))$ to minimize the unconstrained mean achievable distortion $(\overline{D}(I_t))$ when the transmit power constraint is $P_t$. Mathematically speaking, we are going to address the following optimization problem,

$$\overline{D}(I_t) = \min_{I_t(.)} \int_0^\infty f_t(\gamma) \mathcal{G}(D_r(\gamma)) d\gamma$$

$$\text{s.t.} \begin{cases} P(I_t) = \int_0^\infty \frac{I_t(\gamma)}{\gamma^2} d\gamma \leq P_t \\ I'_t(\gamma) \geq 0 \end{cases} \tag{24}$$

where $f_t(.)$ is the probability density function associated with the first hop. Similar to what is done in the previous subsection, we assume that the last constraint is met, thus considering a single power allocation interval and employing a variational form of the integrands in (24) as $\mathcal{H}(I_t, I'_t, \gamma) = f(\gamma) \mathcal{G}(D_r(\gamma))$ and $\mathcal{W}(I_t, I'_t, \gamma) = \frac{I_t(\gamma)}{\gamma^2}$, one may re-write (24) as,

$$\begin{aligned}\overline{D}(I_t) &= \int_{\gamma_1}^{\gamma_2} \mathcal{H}(I_t, I'_t, \gamma) d\gamma + F_t(\gamma_1) \mathcal{G}(D_r(\gamma_1)) \\ &+ (1 - F_t(\gamma_2)) \mathcal{G}(D_r(\gamma_2)),\end{aligned} \tag{25}$$

$$P(I_t) = \int_{\gamma_1}^{\gamma_2} \mathcal{W}(I_t, I'_t, \gamma) d\gamma + \frac{I_t(\gamma_2)}{\gamma_2} - \frac{1}{\gamma_1} \leq P_t. \tag{26}$$

Similar to what is done in Subsection III-A, the lagrangian functions associated with (25) and (26) can be embedded as $\mathcal{L}(I_t) = \overline{D}(I_t) + \lambda\big(P(I_t) - P_t\big)$. Taking an arbitrary increment like $\delta I_t(\gamma)$ on the auxiliary function $I_t(\gamma)$ and equating the linear part of $\Delta(\delta I_t) = \mathcal{L}(I_t + \delta I_t) - \mathcal{L}(I_t)$ to zero, yields,

$$\int_{\gamma_1}^{\gamma_2} \big[\mathcal{H}_{I_t} + \lambda \mathcal{W}_{I_t} - \frac{d}{d\gamma}[\mathcal{H}_{I_t'} + \lambda \mathcal{W}_{I_t'}]\big]\delta I_t(\gamma)d\gamma$$
$$+\big(\mathcal{H}_{I_t'} + \lambda \mathcal{W}_{I_t'}\big)\Big|_{\gamma=\gamma_2} \delta I_t(\gamma_2) +$$
$$\bigg(\big(1 - F_t(\gamma_2)\big)\mathcal{G}_{D_r}\big(D_r(\gamma_2)\big)\frac{\partial D_r(\gamma_2)}{\partial I_t(\gamma_2)} + \frac{\lambda}{\gamma_2}\bigg)\delta I_t(\gamma_2) = 0, \tag{27}$$

where the term $\mathcal{G}_{D_r}\big(D_r(\gamma)\big)$ represents the partial derivation of $\mathcal{G}\big(D_r(\gamma)\big)$ with respect to $D_r(\gamma)$, i.e., $\mathcal{G}_{D_r}\big(D_r(\gamma)\big) = \frac{\partial \mathcal{G}\big(D_r(\gamma)\big)}{\partial D_r(\gamma)}$. As a result, equating the integrand of (27) to zero gives,

$$I_t(\gamma) = \bigg(\frac{b\gamma^2 f_t(\gamma)\mathcal{G}_{D_r}\big(D_r(\gamma)\big)}{\lambda}\bigg)^{\frac{1}{b+1}} \tag{28}$$

Using $I_t(\gamma_1) = 1$, (28) converts to,

$$I_t(\gamma) = \bigg(\frac{\gamma^2 f_t(\gamma)\mathcal{G}_{D_r}\big(D_r(\gamma)\big)}{\gamma_1^2 f_t(\gamma_1)\mathcal{G}_{D_r}\big(D_r(\gamma_1)\big)}\bigg)^{\frac{1}{b+1}}. \tag{29}$$

Noting $D_r(\gamma) = I_t(\gamma)^{-b}$, equating the non-integral term of (27) to zero, gives,

$$\big(1 - F_t(\gamma_2)\big)\mathcal{G}_{D_r}\big(D_r(\gamma_2)\big)\frac{-b}{I_t(\gamma_2)^{b+1}} + \frac{\lambda}{\gamma_2^2} = 0. \tag{30}$$

Extracting $\lambda$ from (28) and (30) and after some mathematics, one can arrive at the following through setting $\gamma = \gamma_2$,

$$\gamma_2 f_t(\gamma_2) = 1 - F_t(\gamma_2). \tag{31}$$

For instance, (31) leads to $\gamma_2 = 1$ in Rayleigh block-fading case. To find the optimum value of mean achievable distortion, $\gamma_2$ is first calculated from (31). Using $\gamma_2$, one can calculate $\gamma_1$ from (26). Thus, $I_t(\gamma)$ can be derived from (29). Then $D_r(\gamma)$ is computed from $D_r(\gamma) = I_t(\gamma)^{-b}$. Finally, the optimum value of the average achievable distortion at the destination can be computed using (25).

It should be noted that the material provided in this section is based on the assumption that $\mathcal{G}(.)$ is an increasing and convex function. The numerical results demonstrate the validity of this assumption for a variety of distributions including Rayleigh block-fading channel which is studied in the current work.

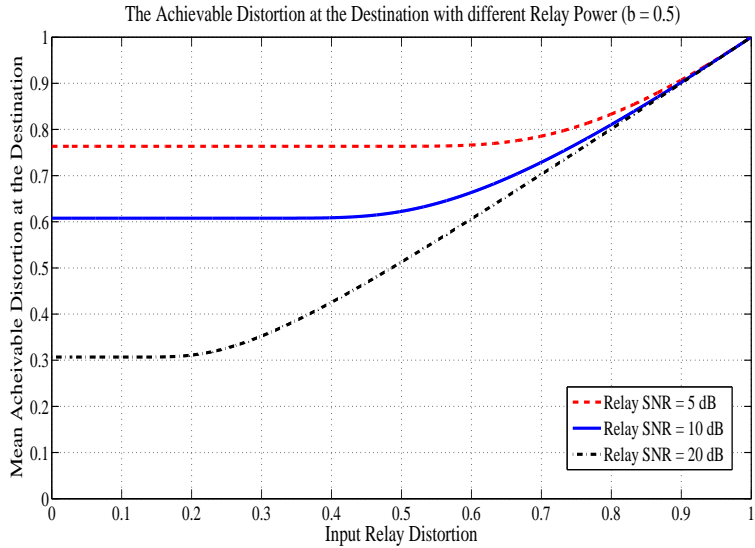

Fig. 1. The effect of relay's transmit power on the mean achievable distortion.

## IV. SIMULATION RESULTS

This section aims at investigating the performance of the proposed method in the terms of the achievable distortion for Rayleigh fading channel.

To find the mean achievable distortion at the destination when the relay is subject to a peak power constraint, the optimal auxiliary function at the relay is derived through using (18) for the interval $l \in [l_1, l_2]$ where $l_1$ and $l_2$ are determined through using equations (11) and (19), assuming the input distortion at the relay is known. Afterwards, the mean achievable distortion at the destination for any given relay's input distortion, i.e., $\mathcal{G}(D_r(\gamma))$, is computed through the use of (9).

For the first hop, inserting the exponential distribution in equation (31) leads to $\gamma_2 = 1$. However, the value of $\gamma_1$ as well as the optimal function $I_t^{opt}(\gamma)$ associated with the first hop have yet to be addressed. It should be noted that $\mathcal{G}(D_r(\gamma))$ can not be analytically derived, hence, $I_t(\gamma)$ can not be analytically extracted from (29). To overcome this issue, since $\mathcal{G}_{D_r}(D_r(\gamma))$ is known for the special case of knowing CSI at the relay [8], we make use of this analytical solution with adjustable power and source/channel mismatch values as two degrees of freedom. This is due to the fact that observed numerical results demonstrate that in a Rayleigh flat fading channel, the function $\mathcal{G}_{D_r}(D_r(\gamma))$ closely follows the shape of [8],

$$\mathcal{G}_{D_r}(D_r(\gamma)) = \exp\left(-\frac{D_r(\gamma)^{\frac{-1}{B}} - 1}{p_r}\right), \tag{32}$$

where $p_r$ and $B$ are degrees of freedoms to be adjusted according to the least square criterion to follow the shape of numerically obtained $\mathcal{G}_{D_r}(D_r(\gamma))$. Thus, noting $\mathcal{G}(D_r(\gamma)) = \int \mathcal{G}_{D_r}(D_r(\gamma))dD_r(\gamma) + Cte$, the function $\mathcal{G}(D_r(\gamma))$ becomes,

$$\mathcal{G}(D_r(\gamma)) = \frac{B}{p_r^B} e^{\frac{1}{p_r}} \Gamma\left(-B, \frac{D_r(\gamma)^{\frac{-1}{B}}}{p_r}\right) + Cte, \tag{33}$$

where the function $\Gamma(a, x)$ has the following definition,

$$\Gamma(a, x) = \int_x^\infty t^{a-1} e^{-t} dt \tag{34}$$

Numerical results indicate that there is a small relative error between the numerical computation of (9) and (33). Using (32), the equivalent auxiliary function $I_t^{opt}(\gamma)$ can be formulated as follows,

$$I_t^{opt}(\gamma) = \left[\frac{b(b+1)p_R W_L\left(\frac{Be^{\frac{B}{b(b+1)p_R}}\left(\frac{\gamma^2 e^{\gamma_1 - \gamma}}{\gamma_1^2}\right)^{\frac{B}{b(b+1)}}}{b(b+1)p_R}\right)}{B}\right]^{\frac{b}{B}}, \tag{35}$$

where $W_L(.)$ is the omega function and is the inverse of the function $f(W) = We^W$. Finally, substituting $I_t^{opt}(\gamma)$ and $\gamma_2 = 1$ into the power constraint (26), gives the value of $\gamma_1$.

For the sake of comparison, the achievable distortion of the proposed method is compared to that of the AF strategy. In AF strategy, it is assumed the relay amplifies the received signal and retransmits it to the destination. In this case, the CDF and pdf of the equivalent channel gain between the source and the destination, namely $s$, when the fading characteristics associated with both hops follow the Rayleigh distribution, can be computed as [11],

$$F_{eq}(s) = 1 - \int_{\frac{P_t}{P_r}s}^\infty \exp\left(-l - \frac{s(1 + lP_r)}{lP_r - sP_t}\right) dl, \tag{36}$$

$$f_{eq}(s) = \int_{\frac{P_t}{P_r}s}^\infty \frac{lP_r(1 + lP_r)}{(sP_t - l_P r)^2} \exp\left(-l - \frac{s(1 + lP_r)}{lP_r - sP_t}\right) dl. \tag{37}$$

Thus, the optimal power allocation function can be numerically found through the model proposed in [7]. Fig. 2 is provided to compare the performance of the proposed approach to that of using a single layer code as well as the AF strategy, when the transmit SNR is set to 20dB, while the SNR at the relay varies from 0dB to 30dB.

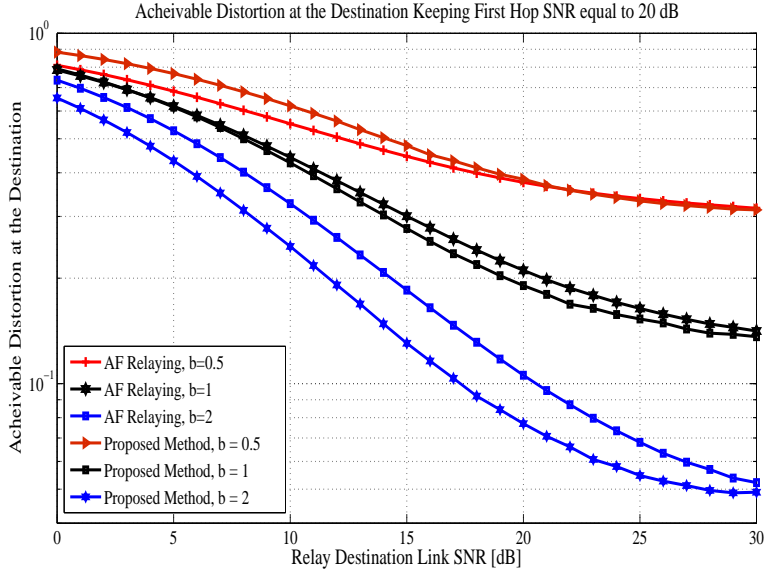

Fig. 2. Comparison results.

## V. Conclusion

This paper aims at investigating the average achievable distortion of a two-hop relay-assisted network employing the decode and forward (DF) strategy assuming the transmitters are unaware of channel gains. In such case, it is argued that multi-layer source coding approach yields better distortion w.r.t. the single-layer code. Accordingly, through using joint successive refinement source coding and multi-layer channel coding approach, the average achievable distortion is related to the power density function across code layers associated with both hops, where the optimal power allocation policies across code layers are numerically derived, leading to the minimum achievable distortion. The result is also compared to that of AF strategy showing for channel mismatch factors greater than unity the DF strategy outperforms the AF at low to moderate SNR regime of the second hop, while at high SNR values the resulting average achievable distortions coincide each other, indicating that since the relays power is relatively high, the first hop is the main bottleneck.

## VI. APPENDIX

### A. Proof of Lemma 1

At first, we are going to state that the the last constraint of (20), results in (21). To this end, the lagrangian function associated with (20) can be written as,

$$\mathcal{L}(I) = \int_0^\infty \left( f(x)\mathcal{G}\left(\frac{1}{I(x)^b}\right) + \lambda \frac{I(x)}{x^2} - I'(x)\phi(x) \right) dx, \qquad (38)$$

where $\lambda$ is the corresponding lagrange multiplier of the first constraint and $\phi(x)$ is an arbitrary positive function, ensuring $I'(x)$ is monotonically non-decreasing at each point (the second constraint). Defining

$$D(x) = \frac{1}{I(x)^b}, \qquad (39)$$

and applying the variational method in [10] to (38), gives the optimal function $I(.)$ as follows,

$$-b\mathcal{G}_D(D(x))\frac{f(x)}{I(x)^{b+1}} + \frac{\lambda}{x^2} + \phi'(x) = 0, \qquad (40)$$

where $\mathcal{G}_D(.)$ denotes the partial derivation with respect to $D(.)$, and $\phi'(x) = \frac{d}{dx}\phi(x)$. Taking the slackness condition into account, in the positive power allocation region, where the second constraint is met, we should have $\phi(x) = 0$. In this case, referring to (40), it follows,

$$I(x) = \left( \frac{bx^2 f(x)\mathcal{G}_D(D(x))}{\lambda} \right)^{\frac{1}{b+1}}. \qquad (41)$$

Here, one should determine the necessary condition for having $I'(x) > 0$. Bringing both sides of (41) to the power of $(b+1)$ and taking derivation w.r.t. $x$, gives,

$$(b+1)I(x)^b I'(x) = \frac{b}{\lambda}\left[ \mathcal{G}_D(D(x))\frac{d}{dx}(x^2 f(x)) + x^2 f(x)\frac{d}{dx}\mathcal{G}_D(D(x)) \right]. \qquad (42)$$

As it is assumed, $\mathcal{G}(.)$ is an increasing convex function, thus, taking partial derivation w.r.t. its input argument, i.e., $\mathcal{G}_D(D(x))$, it results in a positive value. Now, one should go further through the left-most term of (42). The term $x^2 f(x)$ is a non-negative value, as we are dealing with a non-negative $f(.)$ function, so, the term $\frac{d}{dx}\mathcal{G}_D(D(x))$ should be discussed. Incorporating the derivative chain rule, one would have,

$$\frac{d}{dx}\mathcal{G}_D(D(x)) = D'(x)\mathcal{G}_{DD}(D(x)). \qquad (43)$$

Considering the definition of $D(.)$ in (39), the relation (43) converts to,

$$\frac{d}{dx}\mathcal{G}_D(D(x)) = \frac{-b}{I(x)^{b+1}}I'(x)\mathcal{G}_{DD}(D(x)). \qquad (44)$$

Plugging the result of (44) into (42), results in,

$$\left[(b+1)I(x)^b + \frac{b^2}{\lambda I(x)^{b+1}}\mathcal{G}_{DD}(D(x))\right]I'(x) = \frac{b}{\lambda}\mathcal{G}_D(D(x))\frac{d}{dx}(x^2 f(x)). \tag{45}$$

Noting the convexity, as well as the increasing property of $\mathcal{G}(.)$, it can be claimed that the positivity of $I'(x)$ depends on,

$$\frac{d}{dx}(x^2 f(x)) > 0. \tag{46}$$

Now, we are going to state the condition in which there is at most one single interval with positive power allocation, in any region like $x \in [l, u]$ satisfying $\frac{d}{dx}(x^2 f(x)) > 0$. To this end, let's assume the contradiction. For instance, assume there are two disjoint positive power allocation intervals $[x_1, x_2]$ and $[x_3, x_4]$ ($x_1 < x_1 < x_3 < x_4$) in this region with zero power between them, i.e., $I'(x) = 0$ for $x \in (x_2, x_3)$, according to the method of calculus of variations, the corner points $x_c = x_2, x_3$ should satisfy the following equation [10],

$$\mathcal{L}_{I'}\big|_{x=x_c^-} = \mathcal{L}_{I'}\big|_{x=x_c^+} \tag{47}$$

where plugging (38) into (47), we arrive at,

$$\phi(x_c^-) = \phi(x_c^+) \tag{48}$$

According to the slackness condition, the positive power allocation in $[x_1, x_2]$ implies that $\phi(x_2^-) = 0$. Similarly, it follows $\phi(x_3^+) = 0$. Noting this, and referring to (48), it follows,

$$\phi(x_2^-) = \phi(x_2^+) = 0$$
$$\phi(x_3^-) = \phi(x_3^+) = 0 \tag{49}$$

On the other hand, having $I_t'(x) = 0$ within the interval $(x_2, x_3)$ gives the following,

$$I_t(x_2^+) = I_t(x_3^-) \tag{50}$$

Also, from (40), the following holds in the interval $x \in (x_2, x_3)$,

$$\phi'(x) = \mathcal{G}(D_r(x_2))\frac{bf_t(x)}{I_t(x_2)^{b+1}} - \frac{\lambda}{x^2}. \tag{51}$$

This is due to the fact that $D_r(x)$ and $I_t(x)$ are constant in this interval as we have a zero power allocation. Also, noting $\phi'(x_2) = 0$ ($\phi(x) = 0$ for $x \in [x_1, x_2]$), the equation (40) at point $x = x_2$ becomes,

$$\mathcal{G}(D_r(x_2))\frac{bf_t(x_2)}{I_t(x_2)^{b+1}} - \frac{\lambda}{x_2^2} = 0 \tag{52}$$

Replacing $\lambda$ from (52) into (51), we get,

$$\phi'(x) = \frac{b\mathcal{G}(D_r(x_1))}{\gamma^2 I_t(x_1)^{b+1}} \left(x^2 f_t(x) - x_1^2 f_t(x_1)\right). \tag{53}$$

As a result, noting (46) and (53), it follows $\phi'(x) > 0$ in the interval $x \in (x_2, x_3)$, meaning $\phi(x)$ should be an increasing function. This, however, contradicts the corner condition in (49) stating that $\phi(x_2) = \phi(x_3)$. Thus, there should be at most one continues interval in the region which (46) holds.

## B. Proof of the increasing property for $\mathcal{G}(.)$ function

Here, we are going to demonstrate that the function $\mathcal{G}(.)$ is an increasing function. To this end, combining (9) and (18) follows,

$$\begin{aligned}\mathcal{G}(D_r(\gamma)) &= \left(l_1^2 f_r(l_1)\right)^{\frac{b}{b+1}} \int_{l_1}^{l_2} \frac{f_r(l)^{\frac{1}{b+1}}}{l^{\frac{2b}{b+1}}} dl + F_r(l_1) \\ &+ \left(1 - F_r(l_2)\right) D_r(\gamma).\end{aligned} \tag{54}$$

It's worth mentioning that in the (54), the bounds $l_1$ and $l_2$ are indeed functions of $D_r(\gamma)$. One can name the integral of (54), as $\alpha$, i.e.,

$$\alpha = \left(l_1^2 f_r(l_1)\right)^{\frac{b}{b+1}} \int_{l_1}^{l_2} \frac{f_r(l)^{\frac{1}{b+1}}}{l^{\frac{2b}{b+1}}} dl. \tag{55}$$

On the other hand, re-writing the power constraint using (18) gives,

$$\int_{l_1}^{l_2} \left(\frac{l^2 f_r(l)}{l_1^2 f_r(l_1)}\right)^{\frac{1}{b+1}} \frac{1}{l^2} dl + \frac{I_r(l_2)}{l_2} - \frac{1}{l_1} = P_r \rightarrow$$

$$\int_{l_1}^{l_2} \frac{f_r(l)^{\frac{1}{b+1}}}{l^{\frac{2b}{b+1}}} dl = \left(P_r + \frac{1}{l_1} - \frac{D_r(\gamma)^{\frac{-1}{b}}}{l_2}\right) \left(l_1^2 f_r(l_1)\right)^{\frac{1}{b+1}}. \tag{56}$$

Substituting (56) in (55) gives,

$$\alpha = \left(l_1^2 f_r(l_1)\right) \left(P_r + \frac{1}{l_1} - \frac{D_r(\gamma)^{\frac{-1}{b}}}{l_2}\right). \tag{57}$$

Thus, taking partial derivative from (54), considering (57) yields to,

$$\begin{aligned}\frac{\partial \mathcal{G}(D_r(\gamma))}{\partial D_r(\gamma)} &= \frac{\partial(l_1^2 f_r(l_1))}{\partial l_1}\frac{\partial l_1}{\partial D_r(\gamma)}\left(P_r + \frac{1}{l_1} - \frac{D_r(\gamma)^{\frac{-1}{b}}}{l_2}\right) \\ &+ \left(\frac{-1}{l_1^2}\frac{\partial l_1}{\partial D_r(\gamma)} + \frac{D_r(\gamma)^{-\frac{b+1}{b}}}{bl_2}\right. \\ &+ \left.\frac{D_r(\gamma)^{\frac{-1}{b}}}{l_2^2}\frac{\partial l_2}{\partial D_r(\gamma)}\right)(l_1^2 f_r(l_1)) \\ &+ f_r(l_1)\frac{\partial l_1}{\partial D_r(\gamma)} + (1 - F_r(l_2)) \\ &- f_r(l_2)D_r(\gamma)\frac{\partial l_2}{\partial D_r(\gamma)}.\end{aligned}\qquad(58)$$

Re-writing (58) in a better form gives,

$$\begin{aligned}\frac{\partial \mathcal{G}(D_r(\gamma))}{\partial D_r(\gamma)} &= \left[\frac{D_r(\gamma)^{\frac{-1}{b}}}{l_2^2}(l_1^2 f_r(l_1)) - f_r(l_2)D_r(\gamma)\right]\frac{\partial l_2}{\partial D_r(\gamma)} \\ &+ \left[f_r(l_1) - \frac{l_1^2 f_r(l_1)}{l_1^2}\right]\frac{\partial l_1}{\partial D_r(\gamma)} \\ &+ \frac{D_r(\gamma)^{-\frac{b+1}{b}}}{bl_2}(l_1^2 f_r(l_1)) \\ &+ \frac{\partial(l_1^2 f_r(l_1))}{\partial l_1}\frac{\partial l_1}{\partial D_r(\gamma)}\left(P_r + \frac{1}{l_1} - \frac{D_r(\gamma)^{\frac{-1}{b}}}{l_2}\right) \\ &+ (1 - F_r(l_2)).\end{aligned}\qquad(59)$$

According to (19), the first term in (59) is zero. Simplifying the second component shows that, this one is equal to zero, too. Thus (59) changes to,

$$\begin{aligned}\frac{\partial \mathcal{G}(D_r(\gamma))}{\partial D_r(\gamma)} &= \frac{D_r(\gamma)^{-\frac{b+1}{b}}}{bl_2}(l_1^2 f_r(l_1)) + \frac{\partial(l_1^2 f_r(l_1))}{\partial l_1}\frac{\partial l_1}{\partial D_r(\gamma)} \\ &\times \left(P_r + \frac{1}{l_1} - \frac{D_r(\gamma)^{\frac{-1}{b}}}{l_2}\right) + (1 - F_r(l_2)).\end{aligned}\qquad(60)$$

Writing the integral in (56) as $\int_{l_1}^{l_2}\frac{\left(l^2 f_r(l)\right)^{\frac{1}{b+1}}}{l^2}dl$ and taking derivative with respect to $D_r(\gamma)$ yields,

$$\begin{aligned}&\frac{(l_2^2 f_r(l_2))^{\frac{1}{b+1}}}{l_2^2}\frac{\partial l_2}{\partial D_r(\gamma)} - \frac{(l_1^2 f_r(l_1))^{\frac{1}{b+1}}}{l_1^2}\frac{\partial l_1}{\partial D_r(\gamma)} = \\ &\frac{1}{b+1}\frac{\partial(l_1^2 f_r(l_1))}{\partial l_1}\left(P_r + \frac{1}{l_1} - \frac{D_r(\gamma)^{\frac{-1}{b}}}{l_2}\right)\frac{\partial l_1}{\partial D_r(\gamma)} + (l_1^2 f_r(l_1)) \\ &\times\left(\frac{-1}{b}\frac{\partial l_1}{\partial D_r(\gamma)} + \frac{D_r(\gamma)^{\frac{-(b+1)}{b}}}{bl_2} + \frac{D_r(\gamma)^{\frac{-1}{b}}}{l_2^2}\frac{\partial l_2}{\partial D_r(\gamma)}\right).\end{aligned}\qquad(61)$$

Simplifying (61) results,

$$\left[\frac{\left(l_2^2 f_r(l_2)\right)^{\frac{1}{b+1}}}{l_2^2} - \frac{D_r(\gamma)^{\frac{-1}{b}}\left(l_1^2 f_r(l_1)\right)^{\frac{1}{b+1}}}{l_2^2}\right]\frac{\partial l_2}{\partial D_r(\gamma)} =$$
$$\left[\frac{\left(l_1^2 f_r(l_1)\right)^{\frac{1}{b+1}}}{l_1^2} - \frac{\left(l_1^2 f_r(l_1)\right)^{\frac{1}{b+1}}}{l_1^2} + \frac{1}{b+1}\frac{\partial\left(l_1^2 f_r(l_1)\right)}{\partial l_1}\left(l_1^2 f_r(l_1)\right)^{\frac{-b}{b+1}}\right.$$
$$\left.\times\left(P_r + \frac{1}{l_1} - \frac{D_r(\gamma)^{\frac{-1}{b}}}{l_2}\right)\right] + \frac{D_r(\gamma)^{\frac{-(b+1)}{b}}\left(l_1^2 f_r(l_1)\right)^{\frac{1}{b+1}}}{bl_2}. \quad (62)$$

Where simplifying (62) by the use of (19) gives the following relation,

$$\frac{1}{b+1}\frac{\partial\left(l_1^2 f_r(l_1)\right)}{\partial l_1}\left(l_1^2 f_r(l_1)\right)^{\frac{-b}{b+1}}\left(P_r + \frac{1}{l_1} - \frac{D_r(\gamma)^{\frac{-1}{b}}}{l_2}\right)\frac{\partial l_1}{\partial D_r(\gamma)}$$
$$= -\frac{D_r(\gamma)^{\frac{-(b+1)}{b}}\left(l_1^2 f_r(l_1)\right)^{\frac{1}{b+1}}}{bl_2}. \quad (63)$$

Solving the aforementioned equation for $\frac{\partial l_1}{\partial D_r(\gamma)}$ gives,

$$\frac{\partial l_1}{\partial D_r(\gamma)} = -\frac{(b+1)D_r(\gamma)^{\frac{-(b+1)}{b}}\left(l_1^2 f_r(l_1)\right)^{\frac{1}{b+1}}}{bl_2\frac{\partial\left(l_1^2 f_r(l_1)\right)}{\partial l_1}\left(l_1^2 f_r(l_1)\right)^{\frac{-b}{b+1}}\left(P_r + \frac{1}{l_1} - \frac{D_r(\gamma)^{\frac{-1}{b}}}{l_2}\right)}. \quad (64)$$

Applying (64) into (58) and simplifying the relation gives,

$$\frac{\partial \mathcal{G}\left(D_r(\gamma)\right)}{\partial D_r(\gamma)} = -\frac{D_r(\gamma)^{-\left(\frac{b+1}{b}\right)}}{l_2}\left(l_1^2 f_r(l_1)\right) + \left(1 - F_r(l_2)\right), \quad (65)$$

and finally using (19), (65) simplifies to,

$$\frac{\partial \mathcal{G}\left(D_r(\gamma)\right)}{\partial D_r(\gamma)} = 1 - l_2 f_r(l_2) - F_r(l_2). \quad (66)$$

For instance, in Rayleigh fading channel, noting $f_r(l_2) = e^{-l_2}$ and $F_r(l_2) = 1 - e^{-l_2}$, the necessary condition to have $\frac{\partial \mathcal{G}\left(D_r(\gamma)\right)}{\partial D_r(\gamma)} \geq 0$ is $l_2 \leq 1$.